\documentclass[twocolumn,prb,aps,showpacs]{revtex4}

\usepackage{dcolumn}
\usepackage{amsmath}
\usepackage{epsfig}

\newlength{\figwidth}
\preprint{DRAFT}

\begin{document}


\title{Doping dependence of charge-transfer excitations in 
La$_{2-x}$Sr$_x$CuO$_4$}
\author{Young-June Kim}
\author{J. P. Hill}
\affiliation{Department of Physics, Brookhaven National Laboratory,
Upton, New York 11973}
\author{Seiki Komiya}
\author{Yoichi Ando}
\affiliation{Central Research Institute of Electric Power Industry,
2-11-1 Iwado-kita, Komae, Tokyo 201-8511, Japan}
\author{D. Casa}
\author{T. Gog}
\author{C. T. Venkataraman}
\affiliation{CMC-CAT, Advanced Photon Source, Argonne National
Laboratory, Argonne, Illinois 60439}

\date{\today}

\begin{abstract} 

We report a resonant inelastic x-ray scattering (RIXS) study of the doping
dependence of charge-transfer excitations in $\rm La_{2-x}Sr_xCuO_4$. The momentum dependence
of these charge excitations are studied over the whole Brillouin zone in
underdoped (x=0.05) and optimally doped (x=0.17) samples, and compared with
that of the undoped (x=0) sample. We observe a large change in the RIXS
spectra between the x=0 and x=0.17 sample, while the RIXS spectra of the
x=0.05 sample are similar to that of the x=0 sample. The most prominent
effect of doped-holes on the charge excitation spectra is the appearance of
a continuum of intensity, which exhibits a strong momentum-dependence below
2 eV.  For the x=0.17 sample, some of the spectral weight from the
lowest-lying charge-transfer excitation of the undoped compound is
transferred to the continuum intensity below the gap, in agreement with
earlier optical studies. However, the higher energy charge-transfer
excitation carries significant spectral weight even for the x=0.17 sample.
The doping dependence of the dispersion of this charge-transfer excitation
is also discussed and compared with recent theoretical calculations.

\end{abstract}


\maketitle

\section{introduction}
\label{intro}

One of the central questions in the study of correlated electron systems is
how the various types of electronic excitations evolve, as charge carriers
are doped into Mott insulators. Most notably, the essential physics of the
cuprate superconductors is generally regarded as that of a doped Mott
insulator in two dimensions. For the spin degrees of freedom, magnetic
neutron scattering has allowed the exploration of a large part of the phase
space, and a detailed description of the doping dependence of the spin
excitations now exists for many cuprate compounds. In particular, the
incommensurate magnetic excitations in La-based cuprates are well
known.\cite{Birgeneau89,Cheong91,Yamada98} On the other hand, the situation
for charge degrees of freedom is different. Although angle-resolved
photoemission spectroscopy (ARPES) has provided a large amount of
momentum-dependent information on the dispersion of a single
quasiparticle,\cite{Damascelli03} collective charge excitations at nonzero
momentum transfer are much less understood. This is due in large part to
the lack of a suitable experimental probe, except for electron energy loss
spectroscopy (EELS), which suffers from multiple scattering effects at
large momentum transfers. On the other hand, at the zone center, there
exist extensive optical spectroscopy studies of the doping dependence of
the electronic excitation spectrum. For example, as holes are doped into
La$_2$CuO$_4$, Uchida and coworkers\cite{Uchida91} observed that the
spectral weight is transferred from the charge-transfer (CT) excitation to
low-energy excitations. The latter is composed of a Drude peak and a broad
continuum in the mid-infrared (MIR) range, which has yet to be
understood.\cite{Dagotto94}

For a better understanding of the MIR band and other collective
electronic excitations, it is necessary to also obtain information on
the momentum dependence of these excitations.  The resonant inelastic
x-ray scattering (RIXS) technique in the hard x-ray regime has drawn
interest in recent years, because it can provide just such
momentum-dependent information about charge excitations.
\cite{Kao96,Hill98,Abbamonte99,Hasan00,LCO-PRL} In addition, since the
penetration depth of hard x-rays is several microns, RIXS is a bulk
probe, and can also accomodate magnetic fields and high pressures.
Although RIXS probes valence electronic excitations, it has the
additional benefit of element specificity, which is often associated
only with core-level spectroscopies. Recently, we have reported the
observation of two-dimensional (2D)  CT excitations in insulating
La$_2$CuO$_4$,\cite{LCO-PRL} which is the parent compound of hole-doped
La$_{2-x}$Sr$_x$CuO$_4$ (LSCO). In La$_2$CuO$_4$, the lowest energy
excitation has a gap energy of 2.2~eV at the zone center, and exhibits a
large (1~eV) dispersion. A second feature shows a smaller dispersion
(0.5~eV) with a zone-center energy of 3.9~eV. It was argued that these
are both highly-dispersive exciton-like modes, strongly damped by the
presence of the electron-hole continuum. An alternative description of
the observed spectra has recently been proposed by Nomura and Igarashi.
\cite{Nomura04} They attributed the observed two-peak feature to the
fine structure of the Cu $3d$ partial density of states mixed with the O
$2p$ band.

In order to address the question of how these dispersive charge
excitations evolve as the system becomes metallic via hole doping, we have
carried out RIXS experiments on a lightly doped (x=0.05) sample and an
optimally doped sample (x=0.17). As expected from the doping evolution of
optical spectra \cite{Uchida91}, we observe the transfer of spectral
weight from the lowest-lying CT exciton mode to a continuum at low
energies. We also observe that there remains a significant spectral weight
in the CT-like features around 4 eV, even in the x=0.17 sample. However,
the onset energy of the CT excitation shifts to higher energy, and the
low-energy CT exciton mode observed in the undoped sample seems to
disappear in the optimally doped sample.

This paper is organized as follows. In the next section, we describe the
experimental configurations used in the measurements, as well as the
incident energy dependence of the RIXS spectra.  Temperature and momentum
dependence are discussed in Sec.~\ref{sec:result}A and B, respectively. Our
main results, the doping dependence of the low-energy charge excitations,
are presented in Sec.~\ref{sec:doping}.  Finally, we discuss unresolved
issues and possible future experiments in Sec.~\ref{sec:discussion}.

\section{Experimental details}
\label{exp}

The experiments were carried out at the Advanced Photon Source on the
undulator beamline 9ID-B. A double-bounce Si(111) monochromator and a
Si(333) channel-cut secondary monochromator were utilized. A spherical (1
m radius), diced Ge(733) analyzer was used to obtain an overall energy
resolution of 0.4 eV (FWHM).

Our choice of the LSCO system was based on its simple monolayer
structure, and the fact that one can obtain large single crystals over a
wide range of doping levels, which has enabled detailed characterization
of all parts of the temperature-concentration phase 
diagram.\cite{Kastner98}
Although poor surface quality has limited ARPES study of these samples,
a recent series of experiments by Ino and coworkers have provided
a considerable amount of information about the valence band electronic
structure.\cite{Ino97,Ino98,Ino99,Ino00,Ino02} Floating-zone grown single
crystals of La$_{2-x}$Sr$_x$CuO$_4$ with x=0.17 and x=0.05 were used
in our measurements. The as-grown samples were carefully anealed to 
remove
excess oxygen. Their transport properties were reported
previously.\cite{Ando01} The x=0.17 sample has $T_c \approx 42$ K, while
the x=0.05 sample remains insulating down to low temperatures. The (100)
plane of the x=0.17 crystal and the (110) plane of the x=0.05 crystal
was cut and polished for the x-ray measurements. Throughout this paper, we
use the tetragonal notation, with the {\bf a}/{\bf b} direction coinciding
with the Cu-O-Cu bond direction. The scattering plane was vertical and the
polarization of the incident x-ray was kept along the {\bf c}-direction,
that is perpendicular to the copper-oxygen plane. The x=0.17 sample was
mounted on a closed-cycle refrigerator for the low-temperature
measurements, while the x=0.05 sample was mounted on a room-temperature
aluminium sample holder, which was evacuated to reduce the background from
air scattering.

\begin{figure}
\begin{center}
\epsfig{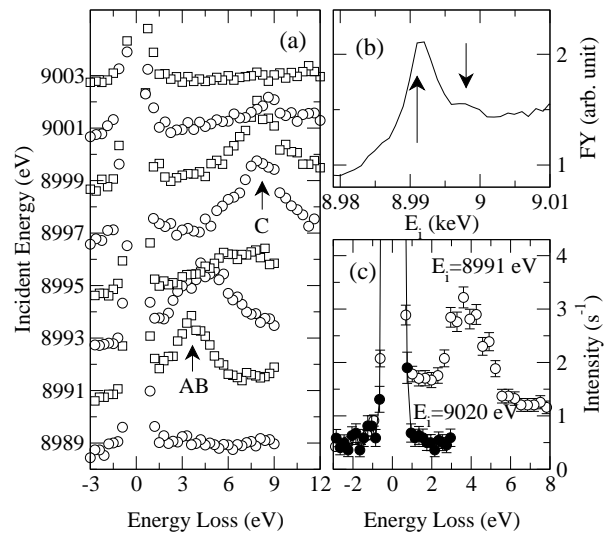}
\end{center}
\caption{(a) The RIXS intensity of the 
x=0.17 
sample is plotted as a 
function 
of energy loss to show incident-energy dependence. The incident energy ($E_i$) of each 
scan is denoted on the left.
(b) X-ray absorption spectrum monitored by fluorescence 
yield. Arrows denote the peak positions, where resonant enhancement are 
observed in the inelastic scattering. (c) Comparison of scattered 
intensity at two different incident energies. At resonance, $E_i=8991$ eV, 
and far from resonance, $E_i=9020$ eV. Data were taken at {\bf Q}={(1 0 
0)}.} 
\label{fig1} 
\end{figure}

In Fig.\ \ref{fig1}(a), we plot the incident energy ($E_i$) dependence of
the RIXS intensity for the x=0.17 sample as a function of energy loss
($\omega=E_i- E_f$) at a fixed momentum transfer of {\bf Q}={(1 0 0)}.  
These data were taken at room temperature. The incident energy of each
scan is denoted on the vertical axis. Since the {\bf Q}={(1 0 0)} position
is a forbidden Bragg peak position, the elastic scattering intenisty
(i.e., at $\omega=0$)  is relatively small, enabling us to study the
low-energy excitations in detail, even at the zone center. Two resonant
features are observed in Fig.~\ref{fig1}(a). As the incident energy is
varied through $E_i \approx 8991$ eV, the low-energy feature around 3~eV
shows a large resonant enhancement in the intensity. In order to compare
the current data with those of the undoped sample,\cite{LCO-PRL} this
excitation is labeled AB. The second feature, labeled C, occurs around
8~eV and shows a markedly different resonance behavior, peaking at $E_i
\approx 8997$ eV. Note that these two resonance energies correspond to
peaks in the x-ray absorption spectra, (arrows in Fig.~ \ref{fig1}(b)).
This latter was obtained by monitoring the fluorescence yield in the same
experimental setup. We find that this resonance behavior for the x=0.17
sample is almost identical to that of undoped La$_2$CuO$_4$ as reported in
Ref.~\onlinecite{LCO-PRL}. Similar resonance behavior was also observed
for x=0.05 (not shown).

In Fig.~\ref{fig1}(c), the $E_i=8991$ eV scan is compared with that
obtained with an incident photon energy of $E_i=9020$~eV, i.e., much
higher than the resonance energy. This latter scan has little, or no
resonant enhancement and represents, in some sense, the background
scattering. From this comparison, one sees that there is significant
resonant inelastic scattering down to 1 eV and lower. This resonant
inelastic scattering is the subject of this paper. In the rest of this
paper, the incident energy is held fixed either at 8991 eV or at 8997
eV, depending on the specific excitation being probed. We can then study
the temperature, momentum, and doping dependence of these charge
excitations in LSCO.

\begin{figure}
\begin{center}
\epsfig{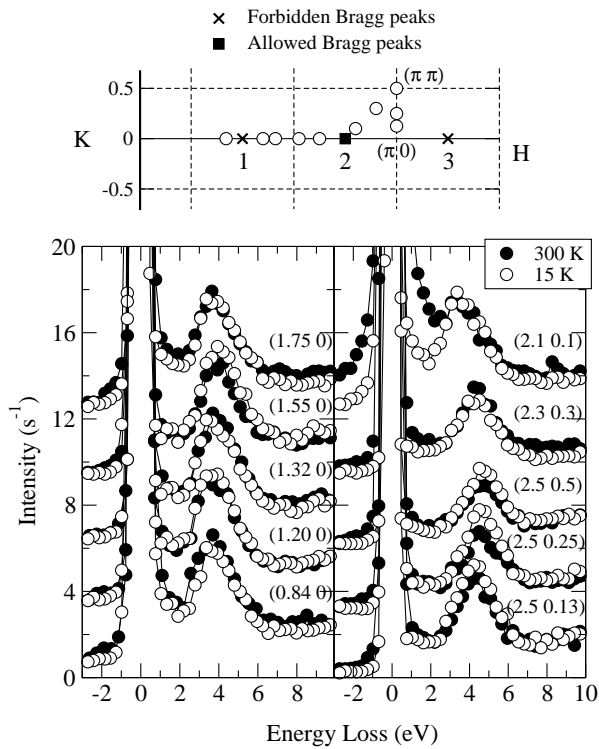}
\end{center}
\caption{RIXS intensity of LSCO (x=0.17) plotted as a function of energy loss at 
various 
momentum transfers. Scans are shifted vertically for clarity. Empty and 
filled symbols are data taken at $T=15$ K and at $T=300$ K, respectively. 
Schematic reciprocal space diagram is shown on top. The {\bf Q}-positions 
where data were collected are denoted as circles.
} \label{fig2}
\end{figure}

\section{experimental results}
\label{sec:result}

\subsection{Temperature Dependence}

In Fig.~\ref{fig2}, representative RIXS scans, taken at various {\bf Q}
positions for the x=0.17 sample, are plotted as a function of $\omega$.  
Here we have fixed the incident energy to 8991 eV, so that we can probe
excitation AB. The out-of-plane component of the momentum transfer is
held constant at $L=0$, and {\bf Q} is denoted for each scan as $(H K)$.  
To note momentum transfer within a single Brillouin zone, we use the
reduced wavevector, ${\bf q} \equiv {\bf Q} - {\bf G}$, where {\bf G} is
a reciprocal lattice vector.  In the left panel, {\bf q} is along the
$(\pi \;0)$ direction, with (1 0)  and (2 0) being zone center
positions. In the right panel, {\bf q} along the $(\pi \;\pi)$ direction
as well as along the zone boundary are shown. These positions are
denoted as open symbols in the reciprocal space diagram shown on top of
Fig.~\ref{fig2}, in which the Brillouin zone boundaries are drawn as
dashed lines.

Let us first consider the temperature dependence. We use filled and empty
symbols in Fig.~\ref{fig2} to denote data taken at $T=300$ K and $T=15$ K,
respectively. Clearly, there is little change in the RIXS spectra between
the two temperatures, which is perhaps not surprising, given that the
energy scale probed in these measurements is very large compared to
thermal energies.\cite{Grenier} As the system is cooled down from room
temperature, the only notable change is the reduced quasi-elastic tail,
which has a large contribution from phonons. This effect is most clearly
seen at the (2.1 0.1)  position, which is close to the {(2 0 0)} Bragg
peak position. This lack of observable temperature dependence in the
charge excitation spectra justifies our later discussion, in which we
compare data taken at different temperatures directly.

\begin{figure}
\begin{center}
\epsfig{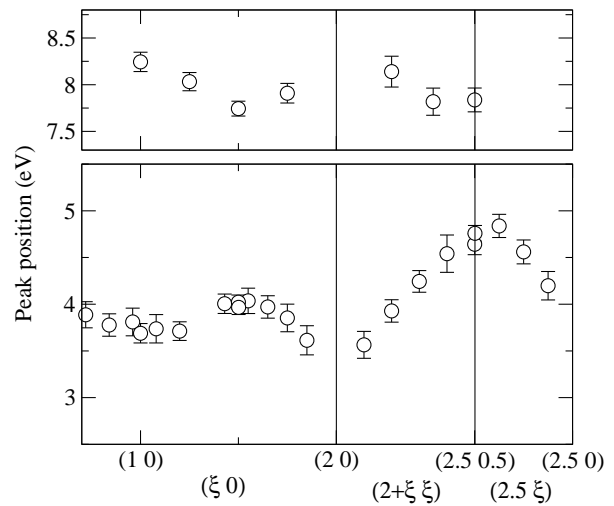}
\end{center}
\caption{Dispersion of the excitations measured at $T=15$ K for LSCO (x-0.17). 
The top 
and bottom panels correspond to the low-energy (AB) and high-energy (C)
excitations 
shown in Fig.~\ref{fig2} and Fig.~\ref{fig4}, respectively.
}
\label{fig3}
\end{figure}

\subsection{Momentum Dependence}

The scans shown in Fig.~\ref{fig2} exhibit quite a large dispersion along
the $(\pi \;\pi)$ direction, while the dispersion along the $(\pi \;0)$
direction is smaller. We have fitted each scan to a Lorentzian peak with a
sloping background, to obtain the peak position as a function of momentum
transfer. This is plotted in the lower panel of Fig.~\ref{fig3}, where the
same notation for {\bf Q} as that in Fig.~\ref{fig2} has been used. This
excitation has a direct gap of $\sim 3.7$ eV; that is, the minimum energy
position is the zone center, and the excitation disperses towards higher
energy at the zone boundary: $\sim 4$ eV at $(\pi \;0)$, and $\sim 4.7$ eV
at $(\pi \;\pi)$. One should note that the dispersion displays significant
anisotropy, which is also evident in the zone boundary dispersion
from (2.5 0.5) to (2.5 0). Such a highly-dispersive excitation, AB, is
reminiscent of the results obtained for $\rm La_2CuO_4$, and, in both samples,
originates from a particle-hole pair excitation across the CT gap.
However, there are some differences between the spectra of the two
samples, which will be discussed in the next section.

\begin{figure}
\begin{center}
\epsfig{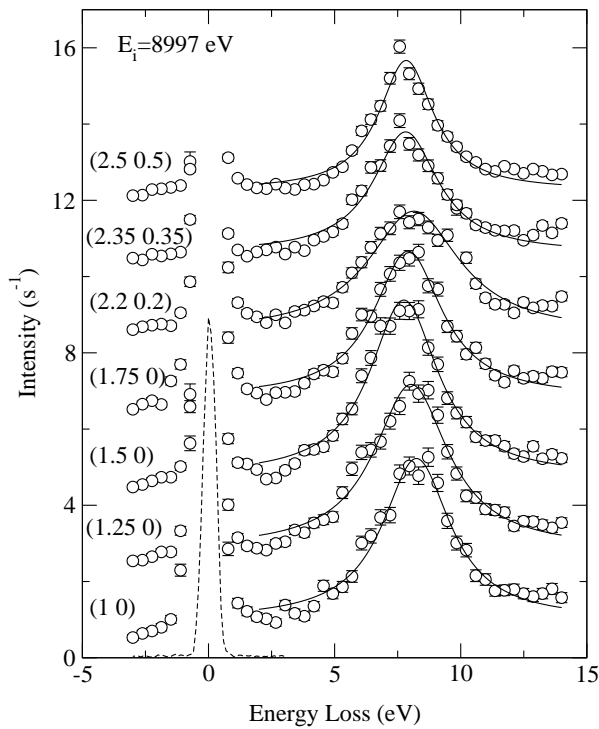}
\end{center}
\caption{RIXS intensity of LSCO obtained at various {\bf Q} with the 
incident 
energy fixed at $E_i=8997$ eV. Data are offset vertically for clarity. The 
dashed line is the intensity of the
elastic scattering scaled to show the experimental energy resolution.}
\label{fig4}
\end{figure}

While excitation AB is resonantly enhanced for incident photon energies
around $E_i=8991$~eV, excitation C resonates around $E_i=8997$~eV, as
discussed above. It is believed to arise from the energy splitting
between the bonding and antibonding molecular orbitals (MO) of the Cu-O
bond.  The observed spectra taken at this incident photon energy at
various momentum transfers are shown in Fig.~\ref{fig4}, with the same
notation for {\bf Q} as in Fig.~\ref{fig2}. Excitation C is intrinsically
very broad, when compared to the energy resolution (shown as the dashed
line). We have therefore fitted each scan to a single Lorentzian peak,
and obtained the peak position as a function of {\bf Q}. This is plotted
in the upper panel of Fig.~\ref{fig3}. One can readily identify the
finite bandwidth of excitation C: the difference between the excitation
energy measured at (1 0) and (1.5 0) is about $\sim 0.5$ eV. It is
interesting to note that this feature has a minimum energy at the zone
boundary, either at $(\pi, 0)$ or $(\pi,\pi)$. That is, excitation C has
an indirect gap. Previously, we have reported that the excitation energy
between the bonding MO and antibonding MO exhibits a systematic
dependence on the Cu-O bond length.\cite{MO} Since a shorter bond-length
results in stronger Cu $3d$ and O $2p$ hybridization, we expect the
bonding-antibonding splitting to grow larger as the doping increases and
the bond length decreases.  Specifically, we observe a larger energy
($\sim 8$ eV) for excitation C in LSCO (x=0.17) than in the x=0 sample,
where an energy of $\sim 7.3$~eV was observed.\cite{MO} We attribute this
to the shorter Cu-O bond length in the x=0.17 sample.

\begin{figure}
\begin{center}
\epsfig{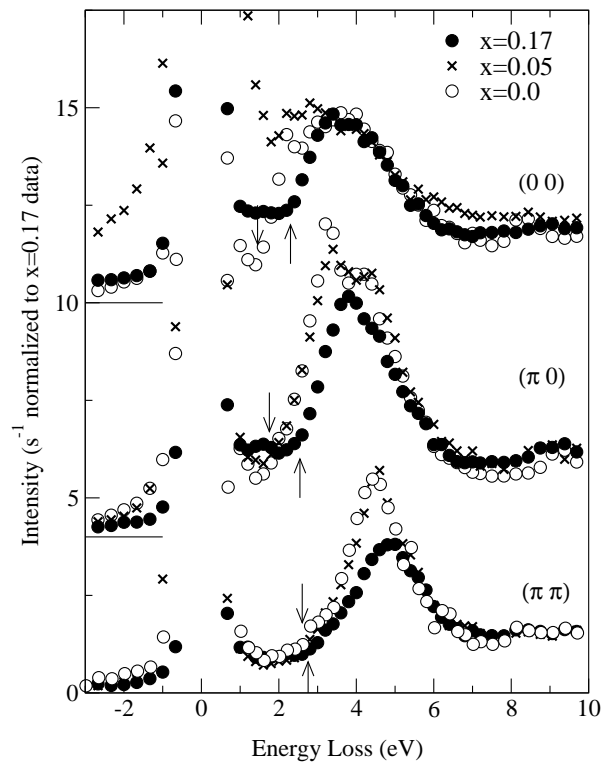}
\end{center}
\caption{Comparison of RIXS spectra at selected momenta for the x=0, 
x=0.05, 
and x=0.17 samples of $\rm La_{2-x}Sr_xCuO_4$. The 
(0 0) and ($\pi$ 0) data are shifted for clarity. The 
up and down arrows denote the onset energy of the spectra for 
the x=0.17 and x=0 samples, respectively.} 
\label{fig5}
\end{figure}

\subsection{Doping Dependence}
\label{sec:doping}

This section reports the main result of this paper. In Fig.~\ref{fig5},
the RIXS spectra of the x=0, x=0.05, and x=0.17 samples are compared at
three high-symmetry {\bf q}-positions: $(0 \;0)$, $(\pi \;0)$, and $(\pi
\;  \pi)$. The data for the x=0 sample were previously reported in
Ref.~\onlinecite{LCO-PRL}. The absolute {\bf Q}-positions are (1 0 0),
(2.5 0 0), and (2.5 0.5 0), from top to bottom for the x=0 and x=0.17
samples. For the x=0.05 sample, the {\bf Q}-positions correspond to (2 1
0), (2 1.5 0), and (1.5 1.5 0), respectively, since the surface of this
sample was cut perpendicular to the [1 1 0] direction. In order to
normalize the intensity between different samples, and between the
spectra obtained at different momentum transfers, we monitored the
fluorescence yield by measuring the K$\beta_5$ emission line at
$E_f=8973$ eV. The fluorescence yield was taken to be a measure of the
volume of the sample being probed and any changes in this were used as
normalization factors to scale the associated inelastic spectra. At {\bf
q}=(0 0) for the x=0 and x=0.17 samples, full emission spectra were
measured and used for the normalization purpose, while for the other
scans shown in Fig.~\ref{fig5}, the intensity at the tail of K$\beta_5$
($E_f \sim 8981$ eV) was used instead. Note that the (0 0) data for
x=0.05 show a large low-energy background intensity from the elastic
tail, since (2 1 0)  is close to the allowed (2 1 1) Bragg reflection.

On a qualitative level, one can compare the {\bf q}=(0 0) data in
Fig.~\ref{fig5} with the optical conductivity data shown in Fig. 7 of
Ref.~\onlinecite{Uchida91}. As La$_2$CuO$_4$ is doped with holes, the
system becomes metallic and the insulating gap of the excitation spectrum
is filled with spectral intensity, arising from the free-carrier
contribution.  However, there is a difference between the doping dependence
of the optical conductivity and the RIXS spectra, especially at low-doping.
In the optical conductivity data of Ref.~\onlinecite{Uchida91}, the gap is
filled even at low doping of x=0.02 and x=0.06. For example, for x=0.02,
the conductivity around $\sim 1$ eV grows quickly to about 60 \% of the
value for x=0.15. On the other hand, although limited in its scope, the
RIXS data in Fig.~\ref{fig5} suggest very little change between the x=0 and
x=0.05 sample, while the higher doping results are consistent with the
optical data. However, one should be
cautious in overintepretating this data, since the relatively poor
signal to noise ratio of the RIXS technique makes it less sensitive to
small changes in spectral weight, when compared to optical conductivity. In
addition, our sample geometry prevented us from obtaining good {\bf q}=$(0
\; 0)$ data for the x=0.05 sample.

\begin{figure}
\begin{center}
\epsfig{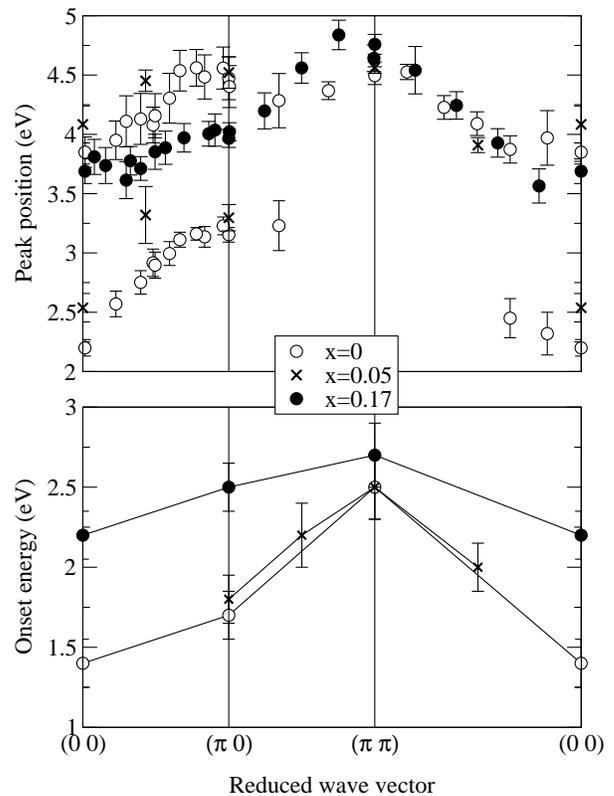}
\end{center}
\caption{Comparison of the dispersion relation of the low-energy charge 
excitation between the $\rm La_{2-x}Sr_xCuO_4$ samples. The top panels show the dispersion of 
peak positions and the bottom panels show the dispersion of onset energies, as 
described in the text.}
\label{fig6}
\end{figure}

For x=0.17, we clearly observe a filling of the gap in the RIXS data,
consistent with the behavior of the optical conductivity.  Figure 1(c)
shows that this gap-filling intensity is much larger than the background
level. We note that the momentum dependence of the continuum intensity is
quite strong. The gap-filling intensity in the region of $\omega < 2$~eV
is significant at $(0 \; 0)$ and at $(\pi \; 0)$, while this gap-filling
intensity becomes very small at $(\pi \; \pi)$. Possible origins of such
momentum-dependent continuum excitations will be discussed in the next
section.

For all doping levels, excitation AB is observed between $2 \sim 5$~eV. For
a quantitative analysis of this scattering, we focus on the doping
dependence of two quantities:  the peak position and the onset energy of
the spectral features. The first quantity can be straightforwardly
determined by fitting the observed spectra to either one or two Lorentzian
peaks. In Ref.~\onlinecite{LCO-PRL}, the x=0 data were fitted with two
peaks, the dispersion of which is reproduced as open circles in
Fig.~\ref{fig6}. Note that the spectral weight of the low energy peak
becomes anomalously small around the $(\pi \; \pi)$, leaving only a single
peak around 4.5 eV. Our attempt to fit the x=0.17 data with two peaks was
unsuccessful, and we have therefore fitted the spectra with a single
Lorentzian with broad width, the peak position of which is plotted as filled 
circles in
Fig.~\ref{fig6}. We also have plotted a limited number of data points for
the x=0.05 sample (crosses). As one can see in Fig.~\ref{fig6}, the
observed excitation of the x=0.17 data more or less falls in the energy
range of the higher-energy excitation of the x=0 sample, which was labeled
B in Ref.~\onlinecite{LCO-PRL}. A possible description, of this
scattering, then, is that excitation A disappears while excitation B
survives, as holes are doped into $\rm La_2CuO_4$.

The onset energy of the spectral features, in contrast to the peak
position, cannot be determined unambiguously, since this requires a
precise knowledge of the background and the lineshape of the spectral
feature. Therefore, we have attempted to determine this quantity consistently in 
the
following way. For the x=0.17 sample, the gap-filling intensity at low
energy is regarded as a constant, and the tail of the spectral feature is
fitted to a linear line. The crossing point of these two linear
extrapolations is taken as the onset energy.  For the x=0 and x=0.05
samples, the minimum intensity is used instead of the constant continuum
intensity as a baseline, and a similar linear extrapolation was used to
determine the onset energy. The onset energies determined in this way are
noted in Fig.~\ref{fig5} (arrows), and are plotted in Fig.~\ref{fig6} as a
function of momentum transfer for different doping levels. One can see
that the onset energies also exhibit some dispersion, albeit smaller than
that of the peak positions.

Upon hole-doping, the onset of excitation AB shifts to higher
energy, with much smaller dispersion than that observed for the undoped
sample. Excitation AB itself becomes broader and shows still sizable
dispersion, especially along the $(0 \; 0)-(\pi \; \pi)$ direction.
However, the relatively sharp and well-defined exciton-like feature (A),
observed around $2 \sim 3$ eV for x=0, is no longer visible in the x=0.17
sample. Since the doped charge-carriers screen the Coulomb interaction
more effectively, it is reasonable to expect that the Coulomb interaction
is no longer strong enough to create even the remnant of the exciton seen
at x=0. Then the excitation AB observed for x=0.17 could be interpreted as
a remainder of the higher energy (B) feature in the x=0 sample, though the
width is broader than that of the undoped sample.

In our previous RIXS study of the undoped sample, we were not able to
identify the nature of the higher energy feature, B. Two possibilities
were suggested: One possibility is that it is also an exciton-like mode,
but with a different symmetry. The second possibility is that feature B
consists mainly of interband transitions from the valence band to
conduction band. Given that the screening presumably prevents the
low-energy exciton mode (A) from binding, our current observation of the
CT feature seems to support the interband transition scenario. This
interpretation is also consistent with the shift of onset energy with
hole doping. Since the chemical potential now lies below the top of the
valence band, a larger onset energy is required to excite a valence
electron to the conduction band.  This behavior of excitation AB is
consistent with the cluster calculation carried out by Tsutsui {\it et
al.} \cite{Tsutsui03} In fact, if one takes the dispersion of the onset
energy, intead of the center of gravity from
Refs.~\onlinecite{Tsutsui03} and \onlinecite{Tsutsui99}, quantitative
agreements are obtained between these calculations and the RIXS results.  
However, this agreement could be fortuitous, as discussed in the next
section.

\section{discussion}
\label{sec:discussion}

There are three main differences observed in our RIXS study of LSCO
(x=0.17 and x=0.05), when compared to the RIXS spectra of undoped $\rm La_2CuO_4$.  
These are i) the appearance of the continuum intensity, ii) changes in the
dispersion and energy scale of excitation AB, and iii) the shift of
excitation C, which is identified as MO excitation. While the origin of
iii) is believed to be mainly structural, the first two are results of
charge carrier doping.

Among these, the low-energy continuum intensity, which fills the
insulating gap, is straightforward to understand. Similar behavior has
been observed with other spectroscopies, such as Raman
scattering\cite{Sugai90} and optical conductivity.\cite{Uchida91} The
gap-filling intensity is believed to arise from the incoherent creation
of particle-hole pairs near the Fermi surface; that is, an intraband
transition across the Fermi level. The momentum dependence of this
gap-filling intensity, however, is difficult to understand.
Specifically, the gap-filling intensity at $(\pi \; \pi)$ is very small
compared to that at zone center, as shown in Fig.~\ref{fig5}.  In their
study of the density-density correlation function of the t-J model,
Tohyama and coworkers also observed similar anisotropic
momentum-dependence of the low-energy spectral weight.\cite{Tohyama95}

One possible source of such anisotropy is the effect of the Fermi
surface topology in intraband transitions. When transfering small
momenta and energy, the transition probability of such an intraband
transition can be quite large, since these transitions involve particle
and holes in the same region of momentum space. However, in order to
transfer large momenta, the particle and hole should come from different
Brillouin zones. In this case, the shape of Fermi surface becomes
important, and the transition probability can become very small. Such an
effect could give rise to the momentum dependence of the very low energy
spectral weight in LSCO. However, it is not clear if this effect can
account for the observed momentum-dependence of the continuum intensity
around $\omega \sim 1$ eV. More measurements with improved energy
resolution are necessary to understand the momentum dependence of the
continuum intensity.

Next, we note that the change in the AB feature is a result of two
distinct phenomena. The first is the shift of chemical potential, which
results in the shift of the AB feature to higher energy. The other is
the disappearance of the lower energy CT exciton mode (mode A),
presumably due to increased charge-carrier screening, and to the
increased number of decay channels. Although our heuristic descriptions
in the previous section provides reasonable accounts of these
observations, a few questions still remain even on the qualitative
level.

One issue is the momentum dependence of the onset energy, shown in
Fig.~\ref{fig6}. Despite the striking resemblance of this figure to the
calculated results in Ref.~\onlinecite{Tsutsui03}, it was pointed out by
Tsutsui {\it et al.} that LSCO should behave differently than the
calculations, because the second and third nearest neighbor hoppings are
much smaller in this material than those used in
Ref.~\onlinecite{Tsutsui03}. In addition, the calculations showed that
the bandwidth of the CT feature decreases in the hole-doped case, while
in the electron-doped case it is almost the same as the undoped case. It
was suggested that the rapid suppression of the antiferromagnetic
correlations in hole-doped LSCO might be the reason for this smaller
dispersion, since the antiferromagnetic correlation remains strong in
electron-doped systems, such as ${\rm Nd_{2-x}Ce_xCuO_4}$. However, one
should note that, even at optimal doping, LSCO exhibits strong
antiferromagnetic fluctuations with a correlation length of $2 \sim 3$
lattice constants.\cite{Birgeneau88} In fact, in underdoped LSCO
systems, static magnetic order (stripes) has been observed at low
temperatures, which implies that antiferromagnetic correlations in the
hole-doped system are not negligible. Thus, if such correlations do play
a role in the dispersion of the charge excitations, as suggested in
Ref.~\onlinecite{Tsutsui03}, then the dispersion in LSCO would be
expected to be closer to the undoped case.

To summarize, we have carried out a resonant inelastic x-ray scattering
(RIXS)  investigation of the doping dependence of charge excitations in
LSCO.  We observe a large change in the RIXS spectra between the x=0 and
x=0.17 sample, while the RIXS spectra of the x=0.05 sample is similar to
that of the x=0 sample. The most prominent effect of the doped-holes is
the appearance of a continuum of intensity below 2 eV, which exhibits a
strong momentum dependence.  For x=0.17, some of the spectral weight
from the charge-transfer excitation of the undoped compound is
transferred to the continuum intensity below the gap, which is
consistent with what has been observed in optical studies. However, the
remainder of the charge-transfer excitation carries significant spectral
weight even for the x=0.17 sample. The observed doping-dependence of the
dispersion of charge-transfer excitations seems to be broadly consistent
with recent theoretical calculations.\cite{Tsutsui03}

\acknowledgements{
We would like to thank A. Fujimori, J. Igarashi, S. Maekawa, T. Nomura, T.  
Tohyama, and K. Tsutsui for invaluable discussions. The work at Brookhaven
was supported by the U. S. Department of Energy, Division of Materials
Science, under contract No. DE-AC02-98CH10886. Use of the Advanced Photon
Source was supported by the U. S. Department of Energy, Basic Energy
Sciences, Office of Science, under contract No.  W-31-109-Eng-38. CAB was
supported by the U. S. Department of Energy, Division of Materials Science,
under contract No. DE-FG02-99ER45772.}

\end{document}